\documentclass{pj}

\usepackage{epsfig}
\usepackage{color}

\usepackage{graphicx}
\usepackage{amsmath}

\def\E{\epsilon}
\def\am{\left(\begin{array}{c}}
\def\amm{\left(\begin{array}{cc}}
\def\a{\end{array}\right)}

\newcommand*{\dya}[1]{{\overline{\overline{\vphantom{#1}\smash{#1}}}}\mathstrut}
\newcommand*{\ve}[1]{{\mathbf #1}}

\newcommand{\bte}{\beta_{\scriptscriptstyle \mathsf{TE}}}
\newcommand{\btm}{\beta_{\scriptscriptstyle \mathsf{TM}}}
\newcommand{\kkk}{\frac{\mathbf k_t \, \mathbf k_t}{k_t^2}}
\newcommand{\nknk}{\frac{\mathbf n \times \mathbf k_t \; \mathbf n \times \mathbf k_t}{k_t^2}}

\begin{document}
\setcounter{page}{1} \pjheader{JEWA (Ikonen {\itshape et al.})}

\title{Vector circuit theory for spatially dispersive uniaxial magneto-dielectric slabs}

\author{Pekka Ikonen, Mikhail Lapine, Igor Nefedov, and Sergei Tretyakov}

\address{Radio Laboratory/SMARAD, Helsinki University of
Technology\\P.O. Box 3000, FI-02015 TKK, Finland.}

\runningauthor{Ikonen} \tocauthor{P.~Ikonen}

\begin{abstract}
We present a general dyadic vector circuit formalism, applicable for uniaxial magneto-dielectric slabs, with
strong spatial dispersion explicitly taken into account. This formalism extends the vector circuit theory,
previously introduced only for isotropic and chiral slabs. Here we assume that the problem geometry imposes
strong spatial dispersion only in the plane, parallel to the slab interfaces. The difference arising from taking
into account spatial dispersion along the normal to the interface is briefly discussed. We derive general dyadic
impedance and admittance matrices, and calculate corresponding transmission and reflection coefficients for
arbitrary plane wave incidence. As a practical example, we consider a metamaterial slab built of conducting
wires and split-ring resonators, and show that neglecting spatial dispersion and uniaxial nature in this
structure leads to dramatic errors in calculation of transmission characteristics.
\end{abstract}

\tableofcontents

\section{Introduction}

It is well known that many problems dealing with reflections from multilayered media can be solved using the
transmission line analogy when the eigen-polarizations are studied separately (see e.g.~\cite{Felsen}). In this
case, the amplitudes of tangential electric and magnetic fields are treated as equivalent \emph{scalar} voltages
and currents in the equivalent transmission line section. In order to account for an arbitrary polarization,
Lindell and Alanen introduced a \emph{vector} transmission-line analogy~\cite{Lindell_Alanen}, where
\emph{vector} tangential electric and magnetic fields serve as equivalent voltage and current quantities. Later
on, the vector transmission-line analogy was further extended for isotropic and chiral slabs into a vector
circuit formalism, with the slabs represented as two-port circuits with equivalent impedances and
admittances~\cite{Oksanen, Analytical}. This vector circuit theory has been successfully applied to study plane
wave reflection from chiral slabs~\cite{Viitanen_IEE1}, and extended to uniaxial multilayer structures
\cite{Serdyukov}.

Recent emergence of metamaterials and the subsequent growth of research interest to their properties,
revitalized the importance of analytical methods for studying artificial media, in particular, for proper
calculation of the reflection and transmission properties. The most prominent examples of
metamaterials~\cite{SPV0} involve split-ring resonators~\cite{PHR99} interlayered with a wire structure, being
organized in an essentially anisotropic manner (e.g., like shown in Fig.~\ref{slab}a). Effective permittivity
(permeability) tensors in such metamaterials correspond to those of uniaxial dielectric (magnetic) crystals, and
the principal values of both tensors differ from unity along one direction only. Moreover, the presence of wire
medium imposes significant spatial dispersion for all waves with an electric field component along the
wires~\cite{Belov-PhysB}.

These peculiarities drive the properties of such metamaterials far apart from what can be described in terms of
the isotropic vector circuit theory, and raise a clear demand for an appropriate generalization. The
corresponding theory extension is the main objective of this paper.

\section{Transmission matrix}

\begin{figure}[b!]
\centering \epsfig{file=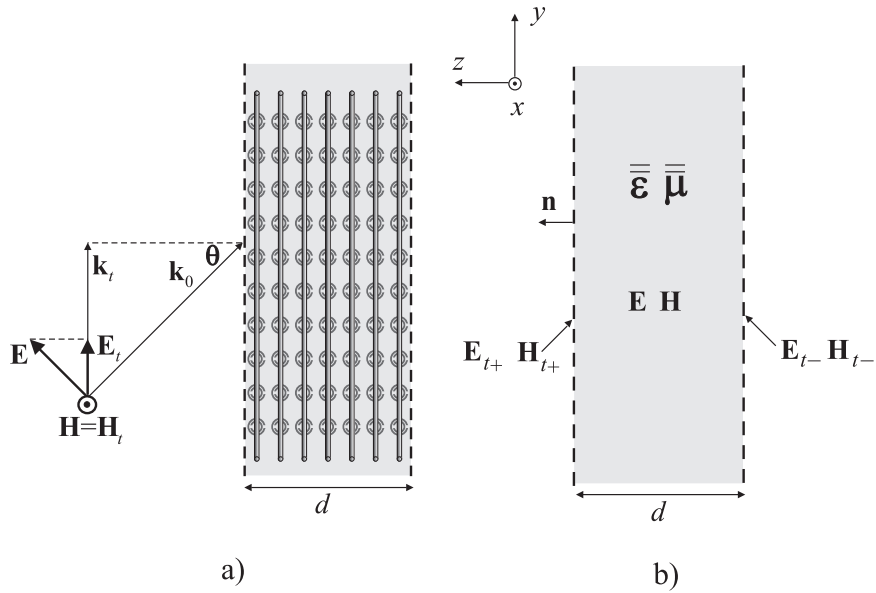, width=11.5cm} \caption{a) A TM-polarized plane wave incident on a slab
implemented using an array of wires and split-ring resonators. b) Macroscopic representation of a uniaxial
magneto-dielectric slab. Subscripts + and $-$ denote the fields at the left and right sides of the slab,
respectively. } \label{slab}
\end{figure}

Let us consider a spatially dispersive slab having thickness $d$ and characterized by the following material
parameter dyadics (Fig.~\ref{slab}b)
\begin{equation}
{\dya{\E}} = \E_t(k_t)
{\dya{I}}_t + \E_n\ve{n}\ve{n}, \quad {\dya{\mu}} = \mu_t
{\dya{I}}_t + \mu_n\ve{n}\ve{n}, \label{param}
\end{equation}
where $\ve{n}$ is the unit normal vector for the slab and ${\dya{I}}_t = {\dya{I}}-\ve{n}\ve{n}$ is the
transversal unit
dyadic.
We consider plane wave excitation and move from the physical space to the Fourier space by a transformation
$\nabla_t\rightarrow-j\ve{k}_t$. For plane waves, $\ve{k}_t$ simply stands for the transversal propagation
factor. Physically we could as well assume that the slab is illuminated by a source located electrically far
away from the slab. Notation $\E(k_t)$ stresses the spatially dispersive nature of the slab in the tangential
plane, and indicates the dependence of the permittivity component from the tangential propagation factor.

\begin{figure}[b!]
\centering \epsfig{file=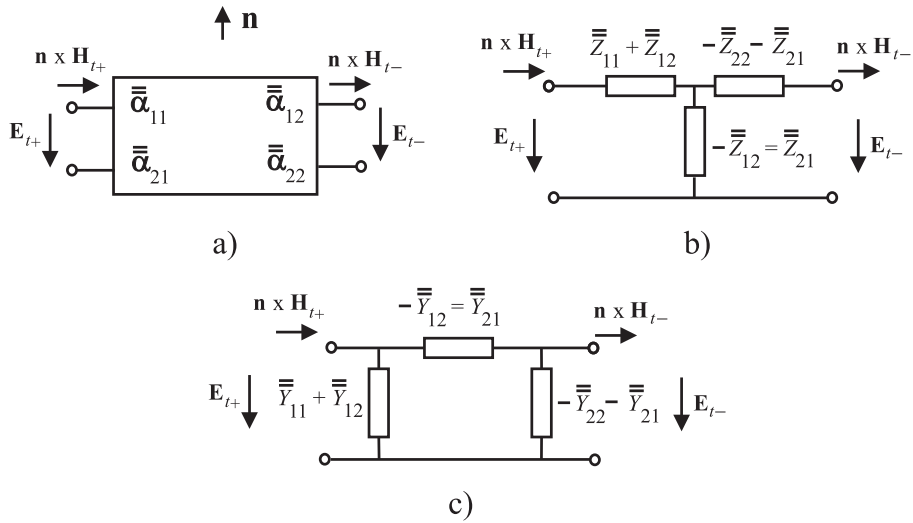, width=12cm} \caption{Different representations for the slab: a) A two-port
transmission line section. b) Vector $\mathsf T$-circuit. c) Vector $\mathsf \Pi$-circuit. The orientation of
the unit vector $\ve{n}$ is in all the cases the same as depicted in a).} \label{circuits}
\end{figure}

Starting from the Maxwell equations the following set of equations can be derived for the tangential field
components
\begin{equation}
\frac{\partial}{\partial{z}}\ve{n}\times\ve{E}_t = -j\omega\mu_t\ve{H}_t
+
\frac{1}{j\omega\E_n}\ve{k}_t\times\ve{k}_t\times\ve{H}_t, \label{rel1}
\end{equation}
\begin{equation}
\frac{\partial}{\partial{z}}\ve{n}\times\ve{H}_t = j\omega\E_t(k_t)\ve{E}_t -
\frac{1}{j\omega\mu_n}\ve{k}_t\times\ve{k}_t\times\ve{E}_t. \label{rel2}
\end{equation}
Next, we integrate eqs.~(\ref{rel1}) and (\ref{rel2}) over $z$ from 0 to $d$:
\begin{equation}
\frac{\ve{n}\times{\ve{E}_{t_+}} - \ve{n}\times{\ve{E}_{t_-}}}{d} =
-j\omega\mu_t{\widehat{\ve{H}}_t}
+\frac{1}{j\omega\E_n}\ve{k}_t\times\ve{k}_t\times{\widehat{\ve{H}}_t}, \label{e1}
\end{equation}
\begin{equation}
\frac{\ve{n}\times{\ve{H}_{t_+}} - \ve{n}\times{\ve{H}_{t_-}}}{d} =
j\omega\E_t(k_t){\widehat{\ve{E}}_t} -
\frac{1}{j\omega\mu_n}\ve{k}_t\times\ve{k}_t\times{\widehat{\ve{E}}_t}. \label{e2}
\end{equation}
Above ${\ve{E}_{t_+}},{\ve{H}_{t_+}}$ refer to the fields at the left side of the slab, and
${\ve{E}_{t_-}},{\ve{H}_{t_-}}$ refer to the fields at the right side of the slab, Fig.~\ref{slab}a. The
averaged fields in eqs.~(\ref{e1}) and (\ref{e2}) are defined as
\begin{equation}
{\widehat{\ve{E}}_t} =
\frac{1}{d}\int_0^d\ve{E}_tdz, \quad {\widehat{\ve{H}}_t} = \frac{1}{d}\int_0^d\ve{H}_t\,dz.
\label{ave}
\end{equation}
After mathematical manipulation (\ref{e1}) and (\ref{e2}) transform into
\begin{equation} \ve{E}_{t_-} -
\ve{E}_{t_+} = -j\omega\mu_td{\dya{A}}\cdot{\ve{n}\times{\widehat{\ve{H}}_t}}
\label{Etrel1}, \end{equation} \begin{equation} \ve{H}_{t_-} - \ve{H}_{t_+} =
j\omega\E_t(k_t)d{\dya{B}}\cdot{\ve{n}\times{\widehat{\ve{E}}_t}}
\label{Htrel1}.
\end{equation}
Above dyadics ${\dya{A}}$ and ${\dya{B}}$ are defined as
\begin{equation}
{\dya{A}} = {\dya{I}}_t - \frac{\ve{k}_t\ve{k}_t}{\omega^2\mu_t\E_n} = \frac{\omega^2\mu_ t\E_n -
k_t^2}{\omega^2\mu_t\E_n} \kkk  + \nknk , \label{A}
\end{equation}
\begin{equation}
{\dya{B}} = {\dya{I}}_t - \frac{\ve{k}_t\ve{k}_t}{\omega^2\mu_n\E_t(k_t)} = \frac{\omega^2\mu_n\E_t(k_t) -
k_t^2}{\omega^2\mu_n\E_t(k_t)} \kkk  + \nknk . \label{B}
\end{equation}
The general solution for the transverse electric field inside the slab reads (now the interest lies only on wave
propagation in $z$-direction)
\begin{equation}
\ve{E}_t(z) = \ve{A}\cdot{e^{-j{\dya{\beta}}{z}}} + \ve
{B}\cdot{
e^{j{\dya{\beta}}{z}}}, \label{Egen}
\end{equation}
where $\ve{A}$ and $\ve{B}$ are
constant vectors
and the $z$-component of propagation factor is different for TM and TE polarizations:
\begin{multline}
{\dya{\beta}} = {\dya{\beta}}(k_t) =  \btm  \kkk  +  \bte  \nknk \\
= \sqrt{\frac{\E_t(k_t)}{\E_n}(\omega^2\mu_t\E_n - k_t^2)}\;  \kkk  +
\sqrt{\frac{\mu_t\vphantom{(k_t)}}{\mu_n}(\omega^2\mu_n\E_t(k_t) - k_t^2)}\; \nknk. \label{betagen}
\end{multline}
Constant vectors $\ve{A}$ and $\ve{B}$ are determined from the boundary conditions
\begin{equation}
\ve{E}(0) = \ve{E}_{t-}, \quad \ve{E}(d) = \ve{E}_{t+}, \label{bc1}
\end{equation}
and the following expressions can be derived
\begin{equation} \ve{A} = (e^{-j\dya{\beta}{d}} - e^{j\dya{\beta}{d}})^{-1}\cdot(\ve{E}_{t+} - \ve{E}_{t-}\cdot e^{j\dya{\beta}{d}}),
\end{equation} \begin{equation} \ve{B} = -(e^{-j\dya{\beta}{d}} - e^{j\dya{\beta}{d}})^{-1}\cdot
(\ve{E}_{t+} - \ve{E}_{t-}\cdot e^{-j\dya{\beta}{d}}). \label{AB}
\end{equation}
After integrating (\ref{Egen}) over $z$ from 0 to $d$ we get for the averaged electric field (note that all the
dyadics are commutative)
\begin{multline}
{\widehat{\ve{E}}_t} =
\frac{1}{jd}(\ve{E}_{t+} + \ve{E}_{t-})\cdot\dya{\beta}^{-1}\cdot(e^{-j\dya{\beta}d} - e^{j\dya{\beta}d})^{-1}
\cdot(2\dya{I}_{t} - e^{-j\dya{\beta}d} - e^{j\dya{\beta}d})\\
= (\ve{E}_{t+} + \ve{E}_{t-})\cdot{\dya{\beta}}^{-1}\cdot\frac{\tan{ \left( \frac{ \dya{\beta} d}{2} \right)}
}{d}. \label{Egenpol2}
\end{multline}
Similarly for the magnetic field \begin{equation} {\widehat{\ve{H}}_t} = (\ve{H}_{t_+} +
\ve{H}_{t_-})\cdot{\dya{\beta}}^{-1}\cdot\frac{\tan \left( \frac{\dya{\beta} d}{2} \right) }{d}. \label{Have1}
\end{equation} Inserting (\ref{Egenpol2}) and (\ref{Have1}) into (\ref{Etrel1}) and (\ref{Htrel1}) leads to the
following result
\begin{equation} \ve{E}_{t_-} - \ve{E}_{t_+} =
-j\omega\mu_td{\dya{A}}\cdot{\ve{n}\times(\ve{H}_{t_+} + \ve{H}_{t_-})}\cdot{\dya{\beta}}^{-1}\cdot\dya{f},
\label{Etrel21}
\end{equation} \begin{equation} \ve{H}_{t_-} - \ve{H}_{t_+} =
j\omega\E_t(k_t)d{\dya{B}}\cdot{\ve{n}\times(\ve{E}_{t_+} + \ve{E}_{t_-})}\cdot{\dya{\beta}}^{-1}\cdot\dya{f}.
\label{Etrel2} \end{equation} Above we have denoted $\dya{f}=\tan({{\dya{\beta}{d}}/2})/d$. Fields at the upper
side of the slab can be expressed with the help of the fields at
the lower side of the slab 
and the following result is obtained after mathematical manipulation: \begin{equation} \ve{E}_{t_+} =
\cos{(\dya{\beta}{d})}\cdot\ve{E}_{t_-} +
j\omega\mu_t\sin{(\dya{\beta}{d})}\cdot{\dya{\beta}}^{-1}\cdot{\dya{A}}\cdot( \ve{n}\times\ve{H}_{t_-}),
\label{f1}
\end{equation} \begin{equation} \ve{n}\times\ve{H}_{t_+} = \cos{(\dya{\beta}{d})}\cdot(\ve{n}\times\ve{H}_{t_-}) +
j\frac{1}{\omega\mu_t}\sin{(\dya{\beta}{d})}\cdot({\dya{\beta}}^{-1}\cdot{\dya{A}} )^{-1}\cdot\ve{E}_{t_-}.
\label{f2}
\end{equation} Writing (\ref{f1}) and (\ref{f2}) into a matrix form we identify the dyadic
transmission matrix for the slab, Fig.~\ref{circuits}a: \begin{equation} \am \ve{E}_{t+} \\
\ve{n}\times{\ve{H}_{t+}} \a = \am {\dya{\alpha}}_{11} \quad
{\dya{\alpha}}_{12}
\\ {\dya{\alpha}}_{21} \quad {\dya{\alpha}}_{22}
\a \cdot \am \ve{E}_{t-} \\
\ve{n}\times{\ve{H}_{t-}} \a, \label{trans_matrix} \end{equation} where the transmission components
read
\begin{gather}
{\dya{\alpha}}_{11} = {\dya{\alpha}}_{22} = \cos({ \btm {d}})
\kkk  + \cos({ \bte {d}}) \nknk , \label{a11}
\\[5pt]
{\dya{\alpha}}_{21} = j\frac{\omega\E_t(k_t)}{ \btm }\sin({ \btm {d}}) \kkk  + j\frac
{ \bte }{\omega\mu_t}\sin({ \bte {d}}) \nknk , \label{a21}
\\[5pt]
{\dya{\alpha}}_{12} = j\frac{ \btm }{\omega\E_t(k_t)}\sin({ \btm {d}}) \kkk  + j\frac
{\omega\mu_t}{ \bte }\sin({ \bte {d}}) \nknk .
\label{a12}
\end{gather}
We immediately notice that if the slab is local and isotropic, coefficients (\ref{a11})--(\ref{a12}) reduce to
those obtained earlier for an isotropic slab \cite{Oksanen}.

The exact boundary condition for a slab on a metal ground plane follows directly from (\ref{f1})
with
$\ve{E}_{t_-}=0$: \begin{equation} \ve{E}_{t_+} = {\dya{Z}}(k_t)\cdot\ve{n}\times
\ve{H}_{t_+},
\end{equation} where the impedance operator reads \begin{equation} {\dya{Z}}(k_t) = j\frac{ \btm }{\omega\E_t(
k_t)}\tan( \btm d) \kkk   +
j\frac{\omega\mu_t}{ \bte }\tan( \bte d) \nknk . \label{imp_op} \end{equation}


\section{Impedance and admittance matrices}

From (\ref{trans_matrix}) it is straightforward to derive the impedance and admittance matrices for the slab \begin{equation} \am
\ve{E}_{t+} \\
\ve{E}_{t-} \a = \am {\dya{Z}}_{11} \quad {\dya{Z}}_{12}
\\ {\dya{Z}}_{21} \quad {\dya{Z}}_{22}
\a \cdot \am \ve{n}\times{\ve{H}_{t+}} \\
\ve{n}\times{\ve{H}_{t-}} \a, \label{imp_matrix} \end{equation}
\begin{equation}
\am \ve{n}\times{\ve{H}_{t+}} \\
\ve{n}\times{\ve{H}_{t-}} \a = \am {\dya{Y}}_{11} \quad
{\dya{Y}}_{12}
\\ {\dya{Y}}_{21} \quad {\dya{Y}}_{22}
\a \cdot \am \ve{E}_{t+} \\
\ve{E}_{t-} \a, \label{adm_matrix}
\end{equation} where the dyadic impedances and admittances depend on
the
transmission components in the following way
\begin{gather}
{\dya{Z}}_{11} =
{\dya{\alpha}}_{11}\cdot{\dya{\alpha}}_{21}^{-1}, \quad
{\dya{Z}}_{12}
=
-{\dya{\alpha}}_{11}\cdot{\dya{\alpha}}_{21}^{-1}\cdot
{\dya{\alpha}}_{22}
+ {\dya{\alpha}}_{12}, \label{Z1}
\\[5pt]
{\dya{Z}}_{21} =
{\dya{\alpha}}_{21}^{-1}, \quad {\dya{Z}}_{22} = -
{\dya{\alpha}}_{21}^{-1}\cdot{\dya{\alpha}}_{22},
\label{Z2}
\end{gather}
\begin{gather}
{\dya{Y}}_{11} =
{\dya{\alpha}}_{22}\cdot{\dya{\alpha}}_{12}^{-1}, \quad
{\dya{Y}}_{12}
= {\dya{\alpha}}_{21} -
{\dya{\alpha}}_{22}\cdot{\dya{\alpha}}_{12}^{-1}\cdot
{\dya{\alpha}}_{11},
\label{Y1}
\\[5pt]
{\dya{Y}}_{21} = {\dya{\alpha}}_{12}^{-1}, \quad
{\dya{Y}}_{22} = -{\dya{\alpha}}_{12}^{-1}\cdot{\dya{
\alpha}}_{11}.
\label{Y2}
\end{gather}
The corresponding $\mathsf T$ and $\mathsf \Pi$-circuit representations are presented in
Fig.~\ref{circuits}b and Fig.~\ref{circuits}c, respectively.

\section{Reflection and transmission dyadics}

Introducing dyadic reflection and transmission coefficients ${\dya{R}}$ and
${\dya{T}}$, eq.~(\ref{trans_matrix}) can be written as two equations in the following
form: \begin{gather}
({\dya{I}}_t + {\dya{R}})\cdot\ve{E}^{\rm inc}_t =
{\dya{\alpha}}_{11}\cdot{\dya{T}} \cdot\ve{E}^{\rm inc}_t +
{\dya{\alpha}}_{12}\cdot{\dya{Z}}_0^{-1}\cdot{\dya{T}}
\cdot\ve{E}^{\rm inc}_t, \label{e11}
\\[5pt]
{\dya{Z}}_0^{-1}\cdot
({\dya{I}}_t -
{\dya{R}})\cdot\ve{E}^{\rm inc}_t =
{\dya{\alpha}}_{21}\cdot{\dya{T}} \cdot\ve{E}^{\rm inc}_t +
{\dya{\alpha}}_{22}\cdot{\dya{Z}}_0^{-1}\cdot{\dya{T}}
\cdot\ve{E}^{\rm inc}_t, \label{e21}
\end{gather}
where $\ve{E}^{\rm inc}$ denotes the incoming electric field and
${\dya{Z}}_0$ is the free space impedance dyadic (seen by the tangential fields)
\begin{equation}
{\dya{Z}}_0 =
\eta_0\cos\theta \kkk  +
\frac{\eta_0}{\cos{\theta}} \nknk . \label{Z0}
\end{equation}
From eqs.~(\ref{e11}) and (\ref{e21}) we can readily solve the transmission and reflection dyadics:
\begin{gather}
{\dya{T}} = 2({\dya{\alpha}}_{11} + {\dya{\alpha}}_{22} +
{\dya{\alpha}}_{12}\cdot{\dya{Z}}_0^{-1} +
{\dya{Z}}_0\cdot{\dya{\alpha}}_{21})^{-1}\cdot{\dya{I}}_t,
\label{T1}
\\[5pt]
{\dya{R}} = ({\dya{\alpha}}_{11} +
\alpha_{12}\cdot{\dya{Z}}_0^{-1})\cdot{\dya{T}} - {\dya{I}}_t.
\end{gather}

\section{Practically realizable slabs}

A typical example which fits into the general model presented above, is a slab of metamaterial, implemented as
an array of conducting wires and split rings resonators (WM--SRR structure), Fig~\ref{slab}a. The wires are
assumed to be infinitely long in the $y$-direction. Moreover, the number of wires in the $x$-direction, and the
number of split-ring resonators both in the $x$- and $y$-directions is assumed to be infinite.

For the case shown in Fig.~\ref{slab}a the non-local permittivity dyadic reads \cite{Belov-PhysB}
\begin{equation} {\dya{\E}} = \E_x\ve{u}_x\ve{u}_x + \E_y\ve{u}_y\ve{u}_y + \E_z\ve{u}_z\ve{u}_z,
\label{e_wires} \end{equation}
\begin{equation}
\E_x=\E_z=
\E_{\rm h},
\quad \E_y = \E_{\rm h} \left( 1 - \frac{k_{\rm p}^2}{k^2 - k_y^2} \right), \label{es}
\end{equation}
where
$\E_{\rm h}$ is
the permittivity of the host matrix, $k_{\rm p}$ is the plasma wave number, $k$ is the wave number
of the host
medium, and $k_y$ is the $y$-component of the wave vector inside the lattice.
Generalization of the dyadic for a case when the wires are perodically arranged along two directions
(double-wire medium with non-connected wires \cite{Nefedov}) is straightforward.

A commonly accepted permeability model as an effective medium description of dense (in terms of the
wavelength) arrays of split-ring
resonators (SRRs) and other similar structures reads (see e.g.~\cite{Kostin,GLS2,Maslovski})
\begin{equation}
{\dya{\mu}} = \mu_x\ve{u}_x\ve{u}_x + \mu_y\ve{u}_y\ve{u}_y +
\mu_z\ve{u}_z\ve{u}_z, \label{mu_rings}
\end{equation}
\begin{equation}
\mu_x = \mu_{\rm h} \left( 1 + \frac{\Lambda\omega^2}{\omega_{\rm 0}^2 - \omega^2 + j\omega\Gamma} \right) ,
\quad \mu_y=\mu_z=\mu_{\rm h}, \label{mu}
\end{equation}
where $\mu_{\rm h}$ is the permeability of the host medium, $\Lambda$ is the amplitude factor ($0<\Lambda<1$),
$\omega_{\rm 0}$ is the undamped angular frequency of the zeroth pole pair (the resonant frequency of the
array), and $\Gamma$ is the loss factor. These parameters can be theoretically estimated for any particular
case~\cite{GLS2}. When the SRRs are positioned to the locations where the quasi-static magnetic field produced
by the wires is zero (to the symmetry planes), there is no near field coupling between the two fractions
\cite{Stas} and the whole metamaterial can be characterized by the permittivity and permeability in the form
above.

The derivation presented above holds for the WM--SRR structure only when the wires are parallel to the slab
interfaces and the tangential propagation factor is restricted to the interval $0\leq{k_t}<k.$ In this case a
plane wave incident on the slab excites only TM and TE wave. If the slab is excited by a source, located close
to the surface, or the source is inside the slab, an additional TEM wave will be excited by the
source,~\cite{Belov-PhysB}. The same happens for an incident plane wave, if $k_t=k$.
Note that, when the wires are perpendicular to the interface different approach is needed.
Recently, authors of
\cite{Pavelarxiv} considered a slab with wires perpendicular to the interfaces and presented another method to
calculate the transmission coefficient for such a slab.

\section{Specific examples}

Here we present some calculated results for plane wave transmission through the WM-SRR slab in Fig.~\ref{slab}b.
We compare the transmission coefficient calculated when the slab is assumed to be uniaxial and non-local with
the transmission coefficient obtained when: (i) the slab is assumed to be uniaxial but local, or (ii) the slab
is assumed to be isotropic and local. Local model for the permittivity means that $k_{y}=0$ in (\ref{es}). Next,
we study the angular dependence of the transmission coefficient at certain frequencies when the effective
refractive index of the slab is close to zero.
For simplicity, only an example of TM polarization is considered. 

For this analysis we assume the following parameters: slab thickness $d=150$ mm,  $\E_{\rm h}=\E_0$, and $k_{\rm
p}=104.7$~m$^{-1}$ (the corresponding plasma frequency is $f_{\rm p}=5$ GHz); $\mu_{\rm h}=\mu_0,\Lambda=0.4,
\omega_0=2\pi\cdot2.5$ GHz, $\Gamma=\omega_0/50$. The permittivity and permeability dyadics are calculated in
accordance with equations \eqref{e_wires}--\eqref{es} and \eqref{mu_rings}--\eqref{mu}, respectively.

\begin{figure}[t!]
\centering \epsfig{file=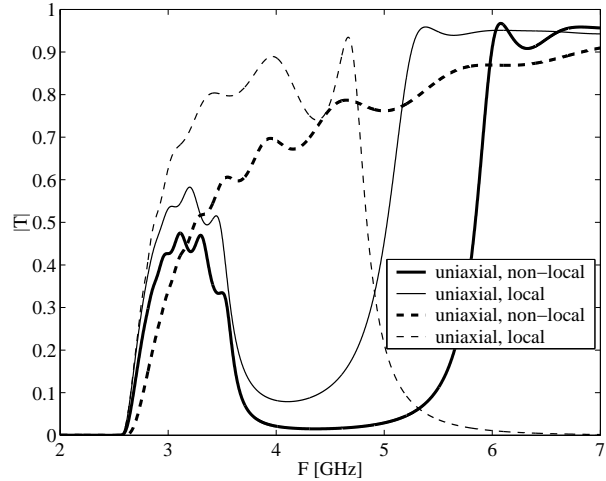, width=8cm} \caption{Transmission coefficient (absolute value) as a function of
frequency at different incidence angles. Exact transmission coefficient (thick lines) is compared to the
transmission coefficient calculated using local permittivity model (thin lines). Solid lines: $\theta=\pi/6$,
dashed lines: $\theta=\pi/3$.} \label{T_ob1}
\end{figure}

Fig.~\ref{T_ob1} compares the exact transmission coefficient to the transmission coefficient calculated when the
slab is assumed to be local [case (i)]. Only for the normal incidence (not plotted) the results are the same. At
small incidence angles (e.g.~$\theta=\pi/6$) there is a transmission maximum around 3 GHz. This maximum
corresponds to the frequency range where both Re$\{\E\}$ and Re$\{\mu\}$ are negative and relatively close to
unity in magnitude. In this situation a planar slab can bring a point source to a focus without spherical
aberration \cite{Pendrylens}. In a certain frequency above 3 GHz Re$\{\mu\}$ becomes positive while Re$\{\E\}$
remains negative, leading to a stop-band. At frequency $f=f_{\rm p}$ permittivity becomes positive and waves can
propagate through the slab. Note that when the permittivity is assumed to be non-local the position of the
pass-band edge is predicted at remarkably higher frequencies compared to the local model.

For larger incidence angles (e.g.~$\theta=\pi/3$) the maxima around 3 GHz disappear and the transmission
coefficient obeys an increasing behavior starting at $f_{\rm 0}$ (the resonant frequency for $\mu_t$). Indeed,
eq.~\eqref{betagen} for $\btm$ shows that the term Re$\{k_0^2(\mu_t - \sin^2\theta)\}$ is negative at
frequencies $f>f_0$.
Thus, a pass-band will appear in the frequency interval $f_0<f<f_{\rm p}$ because in this interval also
Re$\{\E\}$ is negative.

\begin{figure}[b!]
\centering \epsfig{file=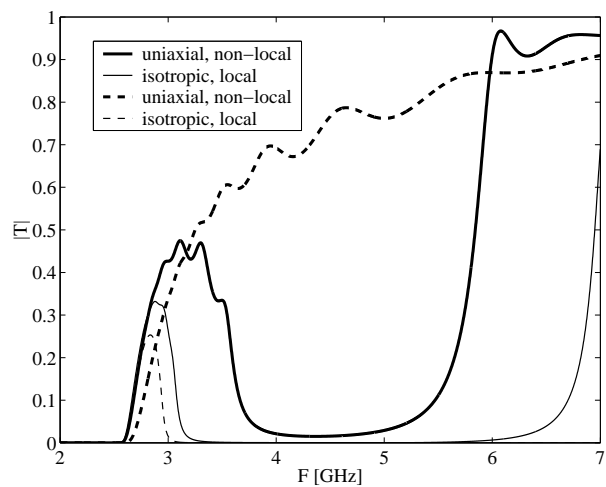, width=8cm} \caption{Transmission coefficient (absolute value) as a function of
frequency at different incidence angles. Exact transmission coefficient (thick lines) is compared to the
transmission coefficient calculated using local and isotropic permittivity model (thin lines). Solid lines:
$\theta=\pi/6$, dashed lines: $\theta=\pi/3$.} \label{T_ob2}
\end{figure}

Fig.~\ref{T_ob2} compares the exact transmission coefficient to the transmission coefficient calculated when the
slab is assumed to be local and isotropic [case (ii)]. Clearly the assumption that the slab is isotropic and
local leads to severe errors except for the normal incidence.


For an additional illustration we study the angular dependence of the transmission coefficient at certain
frequencies when the effective refractive index $n$ of the slab is close to zero while Re$\{\mu_t\}>0$ $\wedge$
Re$\{\mu_t\}\neq\mu_0$, and Re$\{\E_t\}>0$ $\wedge$ Re$\{\E_t\}\neq\E_0$. Comparison is made between the results
given by the non-local and local models (in both models the slab is assumed to be uniaxial). An interest to this
problem arises from recent suggestions for microwave applications benefiting from slabs having low value for
both Re$\{\mu\}$ and Re$\{\E\}$ \cite{Engheta}.

\begin{figure}[t!]
\centering \epsfig{file=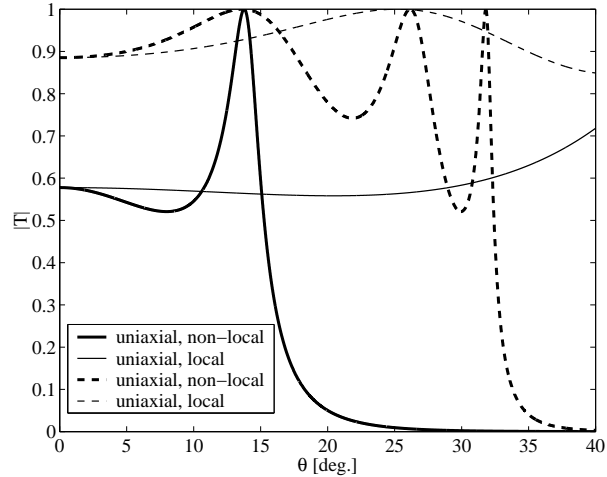, width=8cm} \caption{Transmission coefficient (absolute value) as a function of
the incidence angle at certain frequencies when the slab is assumed to consist only of wires. Exact transmission
coefficient (thick lines) is compared to the transmission coefficient calculated using local permittivity model
(thin lines). Solid lines: $f=1.05\times f_{\rm p}$, dashed lines $f=1.20\times f_{\rm p}$.} \label{th_res2}
\end{figure}

In practise the condition $n\approx 0$, Re$\{\mu_t\}>0$ $\wedge$ Re$\{\E_t\}>0$ occurs at frequencies slightly
above $f_{\rm p}$. We have to bear in mind, however, the physical limitation of the permeability model
(\ref{mu}): the model is valid at low frequencies and at frequencies relatively close to the magnetic resonance.
Here we assume that in the vicinity of $f_{\rm p}$ the permeability model is still valid. The transmission
coefficients are depicted in Fig.~\ref{th_res2}. It shows that transmission maxima, covering a wide range of
angles from 50 to 80 degrees appear when the non-local model is used. These maxima are not predicted when the
local permittivity model is used. This behavior can be explained by considering the $z$-component of the
propagation factor, which in this case can be written in the following form:
\begin{equation} \frac{ \btm ^2}{k_0^2} =  \left( 1 - \frac{k_{\rm p}^2}{k_0^2\cos^2\theta} \right)  \left( \mu_t
- \sin^2\theta \right) . \label{b2} \end{equation} In order for a wave to propagate through the slab, both terms
 inside the parentheses [the right side of eq.~(\ref{b2})] must be simultaneously positive or negative. For small
 incidence angles both
are positive and for large angles both are negative. In a certain range of angles, however, these terms are of
opposite sign, leading to a stop-band.

\begin{figure}[t!]
\centering \epsfig{file=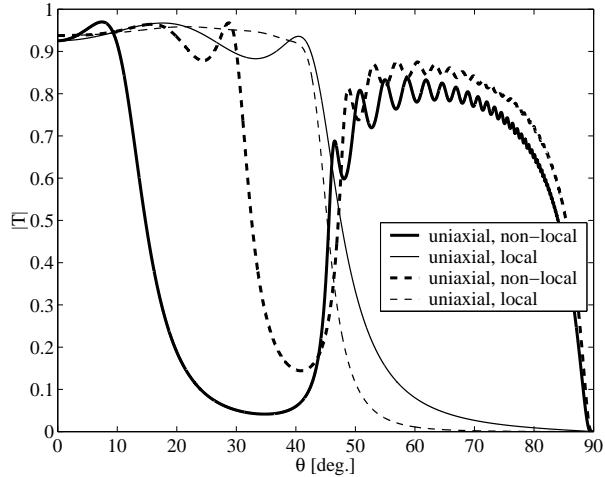, width=8cm} \caption{Transmission coefficient (absolute value) as a function of
the incidence angle at certain frequencies. Exact transmission coefficient (thick lines) is compared to the
transmission coefficient calculated using local permittivity model (thin lines). Solid lines: $f=1.05\times
f_{\rm p}$, dashed lines $f=1.20\times f_{\rm p}$.} \label{th_res1}
\end{figure}

Effectively condition $n\approx0$ can also be achieved in photonic crystals in the vicinity of the stop-band
edge, e.g.~\cite{Gralak, Garcia, Schwartz}. This feature is reported to be important for practical applications
(e.g.~\cite{Enoch}). Accordingly, we can apply the developed method to study the transmission characteristics of
the slab only in the presence of wires. Fig.~\ref{th_res1} shows the calculated results (note the range of
incidence angles). The transmission maxima seen at certain angles correspond to thickness resonances of the
slab. The results indicate that the slab can be utilized as an effective angular filter at microwave
frequencies. Note that when the permittivity is assumed to be local, some maxima are also seen in the
transmission coefficient, however, the location of these maxima is incorrectly predicted.


\section{Conclusions}

In this paper we have formulated a vector circuit representation for spatially dispersive uniaxial
magneto-dielectric slabs.
A dyadic transmission matrix and the corresponding impedance and admittance matrices have been
derived. The results take into account spatial dispersion along the planes parallel to the slab interfaces.

The presented results allow the exact calculation of the transmission and reflection coefficient for a plane
wave with arbitrary incidence angles. This model is applicable, for example, to a typical metamaterial
implemented
as a lattice of conducting wires and split-ring resonators.
It has been shown that for accurate transmission analysis the uniaxial nature
of such a slab, and the spatial dispersion in the wire media must be taken into account.
The calculated results also indicate the feasibility of the slab to
operate as an effective angular filter at microwave frequencies.

\ack

This work has been done within the frame of the European Network of Excellence {\itshape Metamorphose}. The
authors wish to thank Professor~Constantin Simovski and Dr.~Ari Viitanen for stimulating discussions.

\end{document}